\documentclass[aps, pre, reprint]{revtex4-2}
\usepackage{graphicx}
\usepackage[dvipsnames]{xcolor}
\usepackage{amsmath, amssymb}
\usepackage{hyperref}
\usepackage[caption=false]{subfig}

\DeclareMathOperator*{\argmax}{arg\,max} % Jan Hlavacek

\newcommand{\lr}[1]{\left(#1\right)}

\newcommand{\KS}{K-S}

\definecolor{HSafron}{RGB}{232,125,30}

\begin{document}

\title{Learning a Reduced Basis of Dynamical Systems using an Autoencoder}
\author{David Sondak}
\author{Pavlos Protopapas}
\affiliation{Institute for Applied Computational Science, Harvard University}
\date{\today}

\begin{abstract}
  Machine learning models have emerged as powerful tools in physics and engineering. 
  %Although flexible, a fundamental challenge remains on how to connect new machine learning models with known physics. 
  In this work, we present an autoencoder with latent space penalization, which discovers finite dimensional manifolds
underlying the partial differential equations of physics. We test this method on the Kuramoto-Sivashinsky (K-S), Korteweg-de Vries (KdV), and damped KdV equations. We show that the resulting optimal latent space of the K-S equation is consistent with the dimension of the inertial manifold. The results for the KdV equation imply that there is no reduced latent space, which is consistent with the truly infinite dimensional dynamics of the KdV equation. In the case of the damped KdV equation, we find 
  %a negative power law behavior of the 
  that the number of active dimensions decreases with increasing damping coefficient. We then uncover a nonlinear basis representing the manifold of the latent space for the K-S equation. 
  %Penalization of the latent space of an autoencoder could provide novel approaches to building reduced order models for the equations of mathematical physics.
\end{abstract}

\maketitle

\section{Introduction}
Evolution of physical and engineering systems is generally expressed as nonlinear partial differential equations (PDEs). These PDEs are able to capture a wide range of complex phenomena and are therefore indispensable for making predictions of scientific interest. However, most PDEs of practical interest are not analytically tractable. Highly efficient numerical methods can obtain solutions to these PDEs, but are severely limited by the strong multi-scale nature of the underlying dynamics and limitations of hardware resources. There is therefore considerable interest in the development of reduced models that capture only the most important dynamics of the physical phenomenon of interest~\cite{schmid2008dynamic, schmid2010dynamic, benner2015survey, swischuk2019projection}. 
%Indeed, development of reduced and surrogate models is an active field of research with immediate practical implications. 
In recent years, machine learning algorithms have been borrowed from the computer vision community and adapted for physical applications~\cite{lagaris1998artificial, raissi2017physics, wu2018physics, raissi2019physics, greydanus2019hamiltonian, brenner2019perspective, qian2020lift, mattheakis2020hamiltonian, page2020revealing}, offering an enticing approach for blending data with physical principles. In fact, a key goal in merging machine learning with physics problems is to embed known physical laws into machine learning algorithms~\cite{raissi2019physics}. The current work presents a general approach to finding a dynamically-relevant manifold of canonical PDEs using autoencoders. Other recent work has applied autoencoders to learning inertial manifolds of PDEs in a physically-meaningful manner~\cite{gonzalez2018deep, kuptsov2019estimating, champion2019data, linot2020deep, gilpin2020deep}. In particular,~\cite{linot2020deep} 
%follows a similar approach to the method presented here, in which the latent space of an autoencoder is regularized to find the dimension of the inertial manifold.
introduces the Hybrid Neural Network (HNN), in which an autoencoder is used to learn the difference between the data and a
linear projection onto the principle component analysis (PCA) basis. The HNN embeds translation invariance and energy
conservation and learns the dynamics on the inertial manifold of the Kuramoto-Sivashinsky equation. In the present work, the
traditional mean-squared error loss function between the input and reconstruction of the autoencoder is augmented with the
sparsity-promoting mean absolute error loss function, which is applied to the latent space. In this way, the latent space of
the trained autoencoder only contains the minimal dimensions needed to reconstruct the solution. This approach is tested on
two equations that are known to have an inertial manifold (the Kuramoto-Sivashinsky equation and the damped Korteweg-de Vries equation) and one equation whose dynamics are truly infinite dimensional (the undamped KdV equation). In the case of the Kuramoto-Sivashinsky equation, the trained autoencoder and latent space is used to find a non-linear reduced solution basis whose dimension is consistent with that of the inertial manifold.

\begin{figure*}
  \subfloat[\label{fig:KS-solutions}]{%
    \includegraphics[width=.49\linewidth]{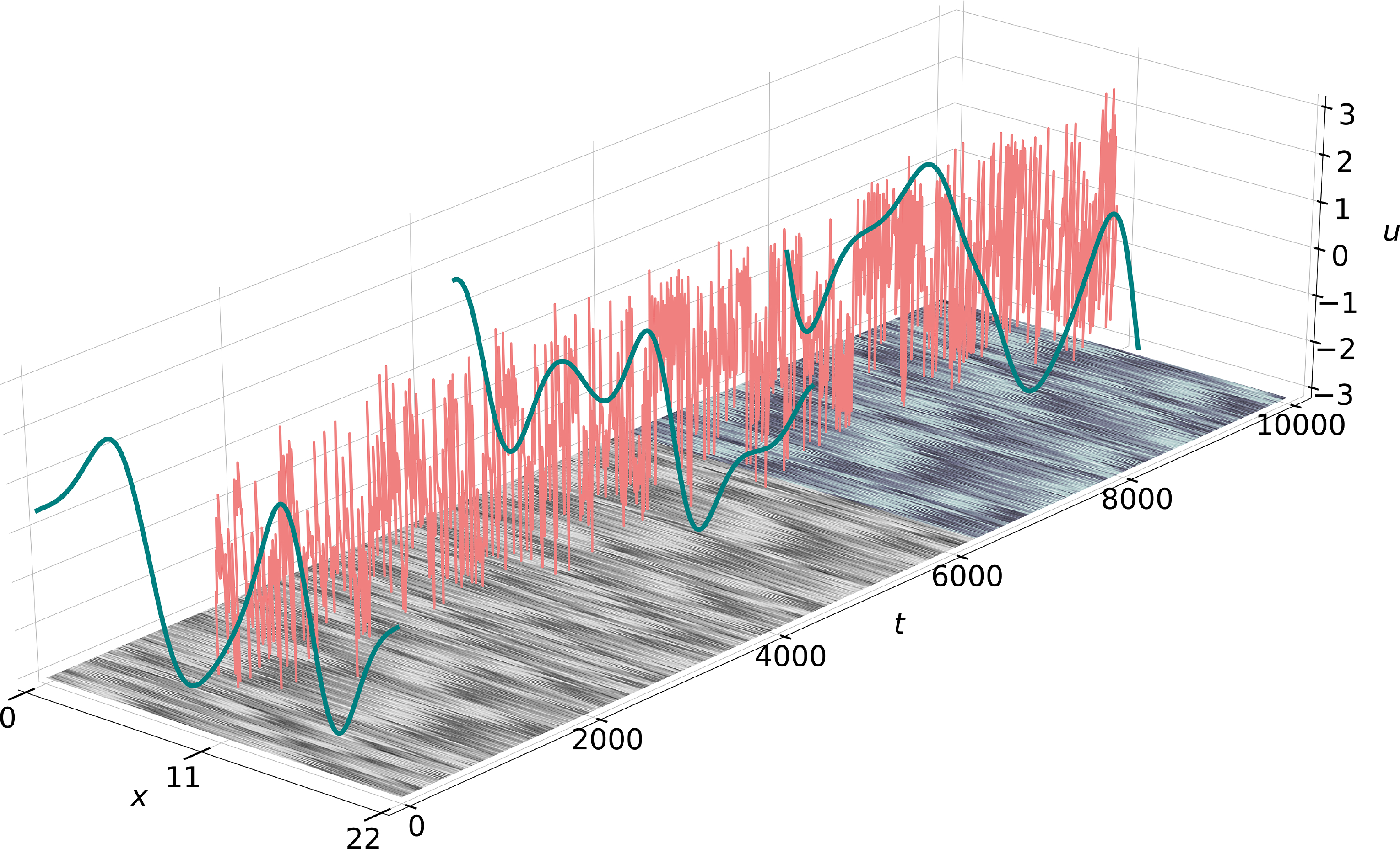}%
  }\hfill
  \subfloat[\label{fig:KdV-solutions}]{%
    \includegraphics[width=.49\linewidth]{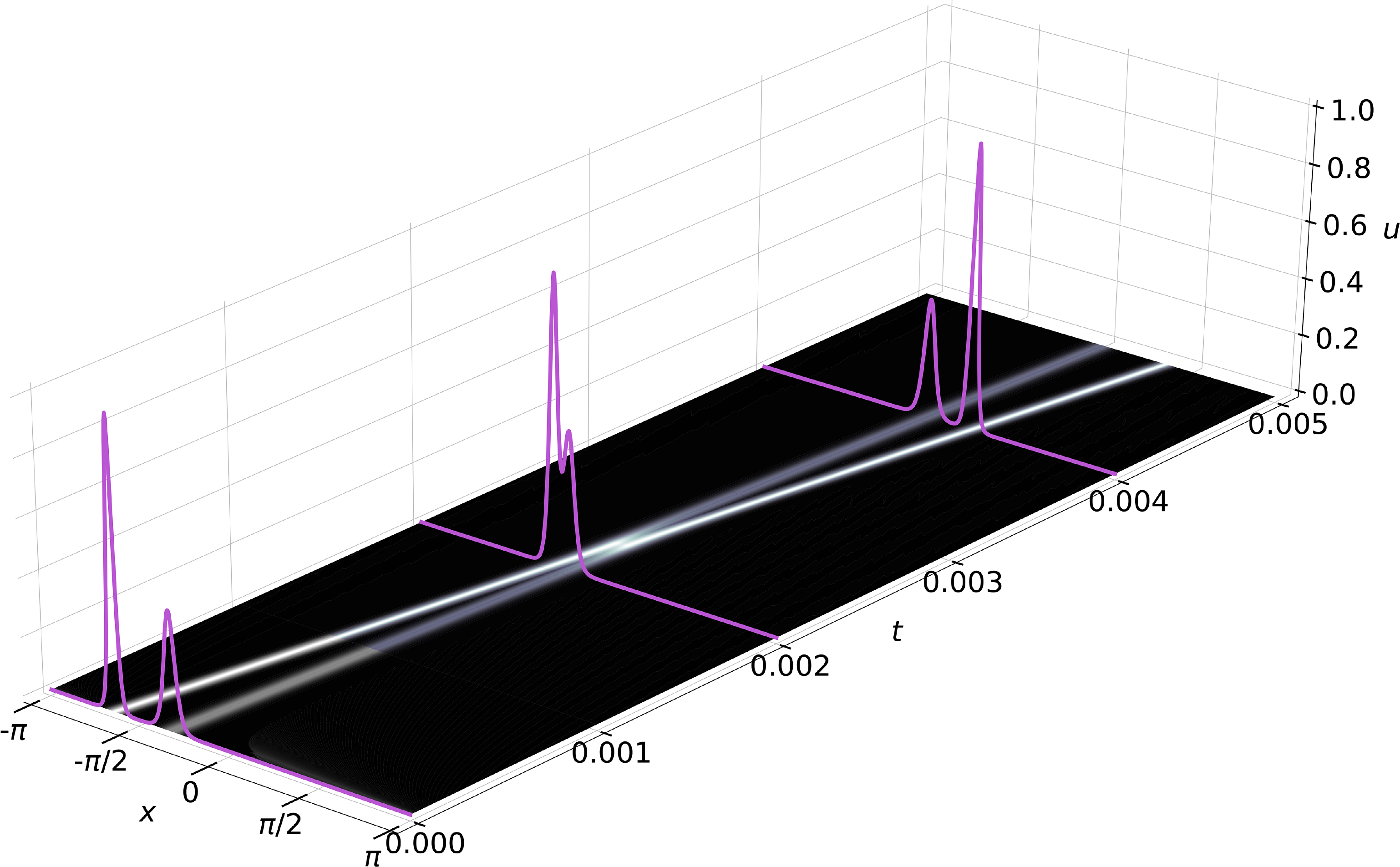}%
  }
  \caption{Datasets from the K-S and KdV equations used in this work. (a) Contours of the K-S equation in space-time with three spatial snapshots at different points in time and a snapshot in time at the center of the domain. The grey shaded area of the contour was not used in the training. (b) Space-time contours of the KdV equation with spatial snapshots in time. The shading is the same as for the K-S equation.}
  \label{fig:pde_solutions}
\end{figure*}

\section{Governing Equations and Datasets}
The Kuramoto-Sivashinsky (\KS) equation is,
\begin{align}
  u_{t} + uu_{x} + u_{xx} + u_{xxxx} = 0 \label{eq:KS}
\end{align}
where $u=u\lr{x,t}$ is the solution field, $x\in\left[0, L\right]$, and $t\in\mathbb{R}^{+}$. The final integration time is denoted by $T$. Equation~\eqref{eq:KS} is subject to periodic boundary conditions $u\lr{0, t} = u\lr{L, t}$ and initial condition $u_{0} = \cos\lr{\frac{2\pi}{L}x}\lr{1 + \sin\lr{\frac{2\pi}{L}x}}$. The \KS\ equation has its roots in physics~\cite{kuramoto1976persistent, cohen1976non, sivashinsky1977nonlinear, sivashinsky1980irregular}, and is a frequently-studied equation in mathematics~\cite{nicolaenko1985some, foias1988inertial, conte1989painleve, jolly1990approximate, robinson1994inertial, jolly2000evaluating, constantin2012integral, ding2016estimating}. Of particular relevance to the present work is the result that the dynamics of the \KS\ equation are confined to an inertial manifold~\cite{foias1988inertial}. That is, despite the \KS\ equation being a non-integrable equation whose solutions exhibit spatio-temporal chaos, the underlying dynamics are exponentially attracted to a finite-dimensional manifold. The dimension of the inertial manifold increases with the bifurcation parameter $L$.

The damped Korteweg–de Vries (KdV) equation, also considered in this work, is,
\begin{align}
  u_{t} + uu_{x} + u_{xxx} - \eta u_{xx} = 0 \label{eq:kdv}
\end{align}
where $x\in\left[-\pi,\pi\right]$ and $\eta\geq 0$ is a damping coefficient. The KdV equation is subject to periodic boundary conditions $u\lr{-\pi, t} = u\lr{\pi, t}$ and initial condition $u_{0} = 3A^{2}\text{sech}^{2}\lr{A\lr{x+2}/2} + 3B^{2}\text{sech}^{2}\lr{B\lr{x+1}/2}$ with $A=16$ and $B=25$. The classical KdV equation is recovered for $\eta=0$. Similarly to the \KS\ equation, the KdV equation is a paradigmatic equation in mathematical physics~\cite{korteweg1895xli, ghidaglia1988weakly}, leading to the discovery of solitons, which are a direct bridge between observed coherent structures in PDEs and nature. The dynamics of the undamped KdV equation are truly infinite dimensional and are therefore not confined to an inertial manifold while those of the damped KdV equation are finite-dimensional~\cite{ghidaglia1988weakly}.

The datasets for this work were generated by solving~\eqref{eq:KS} and~\eqref{eq:kdv} using an exponential time-differencing fourth-order Runge-Kutta method~\cite{kassam2005fourth} and a pseudo-spectral Fourier method in space. 
%Figures~\ref{fig:KS-solutions} and~\ref{fig:KdV-solutions} show the K-S dataset at $L=22$ and the undamped KdV dataset, respectively. 
Spatial snapshots of these solution fields are used as input to the autoencoder. Figure~\ref{fig:pde_solutions} shows examples of the K-S dataset (Figure~\ref{fig:KS-solutions}) and the KdV dataset (Figure~\ref{fig:KdV-solutions}). The numerical parameters used for the cases in this work are presented in Table~\ref{tab:num_sim} in Appendix~\ref{app:num_sum}.

\section{Methodology}
\begin{figure}
    \centering
    \includegraphics[width=0.48\textwidth]{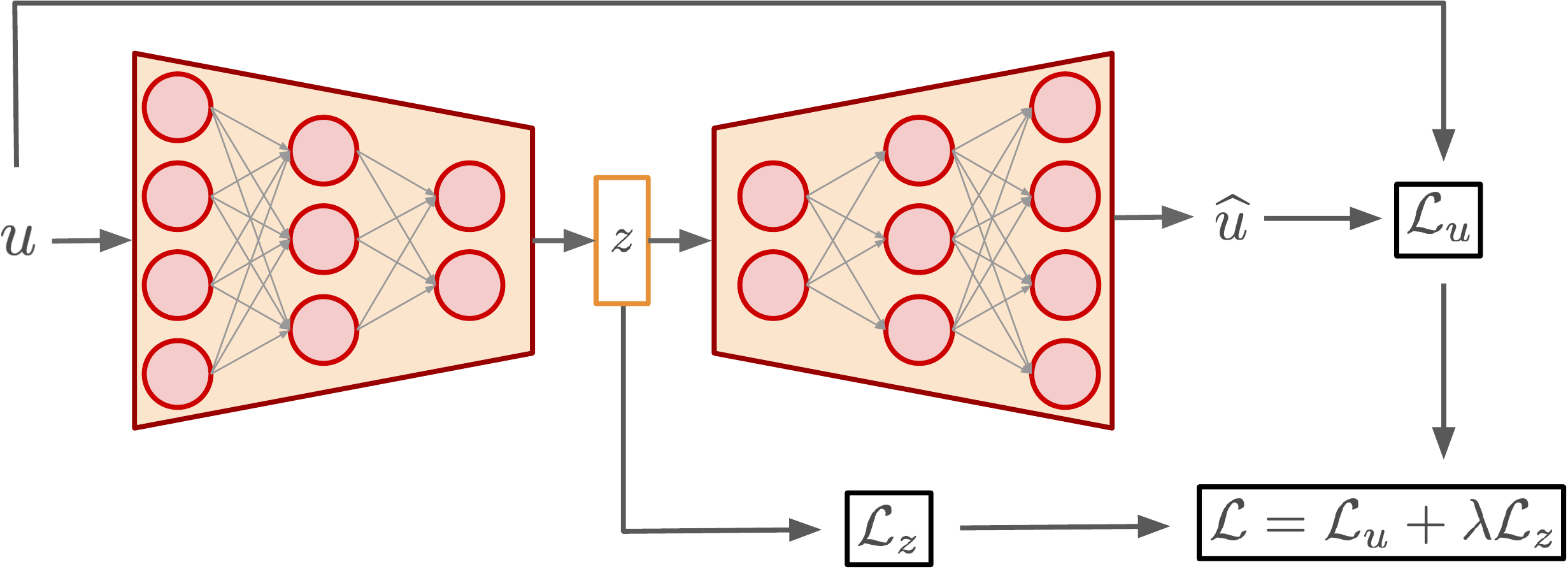}
    \caption{The autoencoder architecture with latent space penalization. The reconstruction loss $L_{u}$~\eqref{eq:MSEloss} is combined with a mean absolute error loss $L_{z}$~\eqref{eq:MAEloss}. The result is a solution reconstruction that uses only relevant latent dimensions.}
    \label{fig:AE}
\end{figure}

Autoencoders are a self-supervised neural network architecture that can be used to find a low-dimensional manifold that represents the data~\cite{rumelhart1985learning, baldi2012autoencoders}. The input to the encoder is mapped to a lower dimensional space called the latent space. The latent space is then expanded through the decoder to reproduce the input to the encoder. An autoencoder with linear activation functions can be shown to be equivalent to the singular value decomposition~\cite{milano2002neural}. In the present work, the input is a snapshot of the solution field obtained from a high-fidelity numerical simulation that used $N$ points in space and $N_{t}$ points in time. A snapshot in space is denoted by $\mathbf{u}_{n} = \mathbf{u}\lr{t_{n}}$ for $n=1,\ldots,N_{t}$ and $\mathbf{u} \in \mathbb{R}^{N}$ is a vector representing the solution at $N$ discrete points in space. This snapshot is mapped to a latent space of dimension $N_{z}$ with $z_{j}\lr{t_{n}}$ the $j^{\text{th}}$ component of the latent space corresponding to snapshot $n$. The reconstructed output of the autoencoder is denoted by $\widehat{\mathbf{u}}_{n}$. The weights and biases associated with each node of the autoencoder are tuned to minimize the total loss,
\begin{align}
  \mathcal{L} = \mathcal{L}_{u} + \lambda \mathcal{L}_{z} \label{eq:loss}
\end{align}
where 
\begin{align}
  \mathcal{L}_{u}\lr{u, \widehat{u}} = \frac{1}{N_{t}N}\sum_{n=1}^{N_{t}}\sum_{i=1}^{N}{\lr{u\lr{x_{i}, t_{n}} - \widehat{u}\lr{x_{i}, t_{n}}}^{2}} \label{eq:MSEloss}
\end{align}
is the mean squared error (MSE) reconstruction loss and 
\begin{align}
  \mathcal{L}_{z} = \frac{1}{N_{t}N_{z}}\sum_{n=1}^{N_{t}}\sum_{j=1}^{N_{z}}{\left|z_{j}\lr{t_{n}}\right|}. \label{eq:MAEloss}
\end{align}
is the mean absolute error (MAE) penalization loss on the latent dimensions. The sparsity of the latent space is controlled by  the regularization parameter $\lambda \geq 0$. A classical autoencoder corresponds to $\lambda=0$. The dimension of the latent space, $N_{z}$, is not known \textit{a priori}, but the MAE penalization promotes sparsity in the latent dimensions while the MSE loss boosts the reconstruction performance. The appropriate value of $\lambda$ will therefore restrict the latent space to the dimensions necessary for a good reconstruction. Figure~\ref{fig:AE} depicts the autoencoder and loss functions used in this work. 
\begin{figure*}
    \centering
    \includegraphics[width=\textwidth]{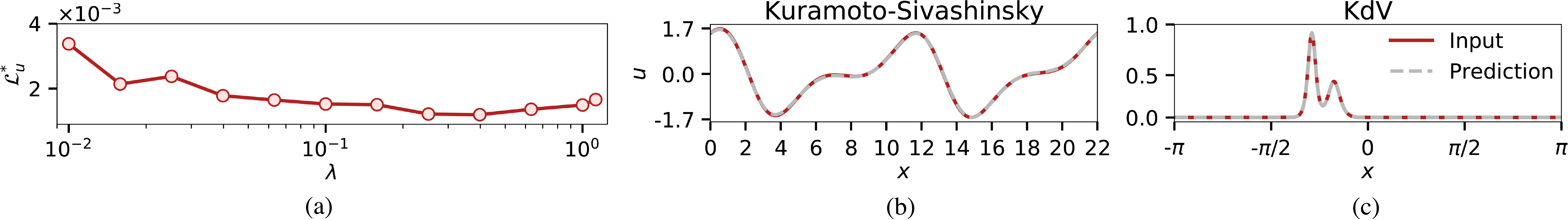}
    \caption{(a) Reconstruction loss on the validation set for the K-S equation with $L=22$ across different values of penalization parameter $\lambda$. In each case the network architecture was $512 \rightarrow 256 \rightarrow 128 \rightarrow 64 \rightarrow 32$. The minimum value occurs at $\lambda = 0.398$. (b) Solution reconstruction and input solution for the K-S equation. (c) Solution reconstruction and input solution for the K-S equation. }
    \label{fig:summary}
\end{figure*}
All networks used in the current work used fully-connected networks with sinusoidal activation functions. For conciseness, we denote %the number of layers and units per layer for the encoder of the network 
the encoder architecture by $N \rightarrow H_{1} \rightarrow H_{2} \cdots \rightarrow H_{D} \rightarrow N_{z}$ where $H_{k}$ represents the number of nodes in layer $k$ and $D$ is the number of hidden layers. The decoder uses the reverse form of the encoder portion. The autoencoders were trained used gradient descent with gradient clipping~\cite{mikolov2012statistical} to limit the maximum gradient to $10$. The Adamax optimizer~\cite{kingma2014adam} was used in all experiments. The transient portion of the dataset was excluded from both the training and validation sets. In general, $80\%$ of the remaining dataset was retained for training and $20\%$ was used for validation. Table~\ref{tab:AE_arch} in Appendix~\ref{app:AE_arch} contains specific details on all autoencoder architectures used in this work including their hyperparameters. 

\section{Results}
Autoencoders were trained on datasets generated from the K-S equation~\eqref{eq:KS} and the KdV equation~\eqref{eq:kdv}. The cases considered in this work are summarized in Tables~\ref{tab:num_sim} and~\ref{tab:AE_arch} in Appendix~\ref{app:num_sum} and~\ref{app:AE_arch}, respectively. Each case was run for a range of regularization parameter values. For each value of $\lambda$, the dataset was split into a training and validation set and the MSE~\eqref{eq:MSEloss} was monitored on the validation set during training. Two models were saved for each experiment. One model was saved at the minimum of the training loss curve while the other model was saved at the minimum of the MSE loss on the validation set. The model with the lowest MSE loss on the validation set at each value of $\lambda$, $\mathcal{L}_{u}^{*}\lr{u_{\text{valid}},\widehat{u}_{\text{valid}}}$, was taken to represent the optimal regularization parameter for that case and was used for analysis.

\subsection{Kuramoto-Sivashinsky Equation}
\begin{figure}
  \centering
  \includegraphics[width=0.48\textwidth]{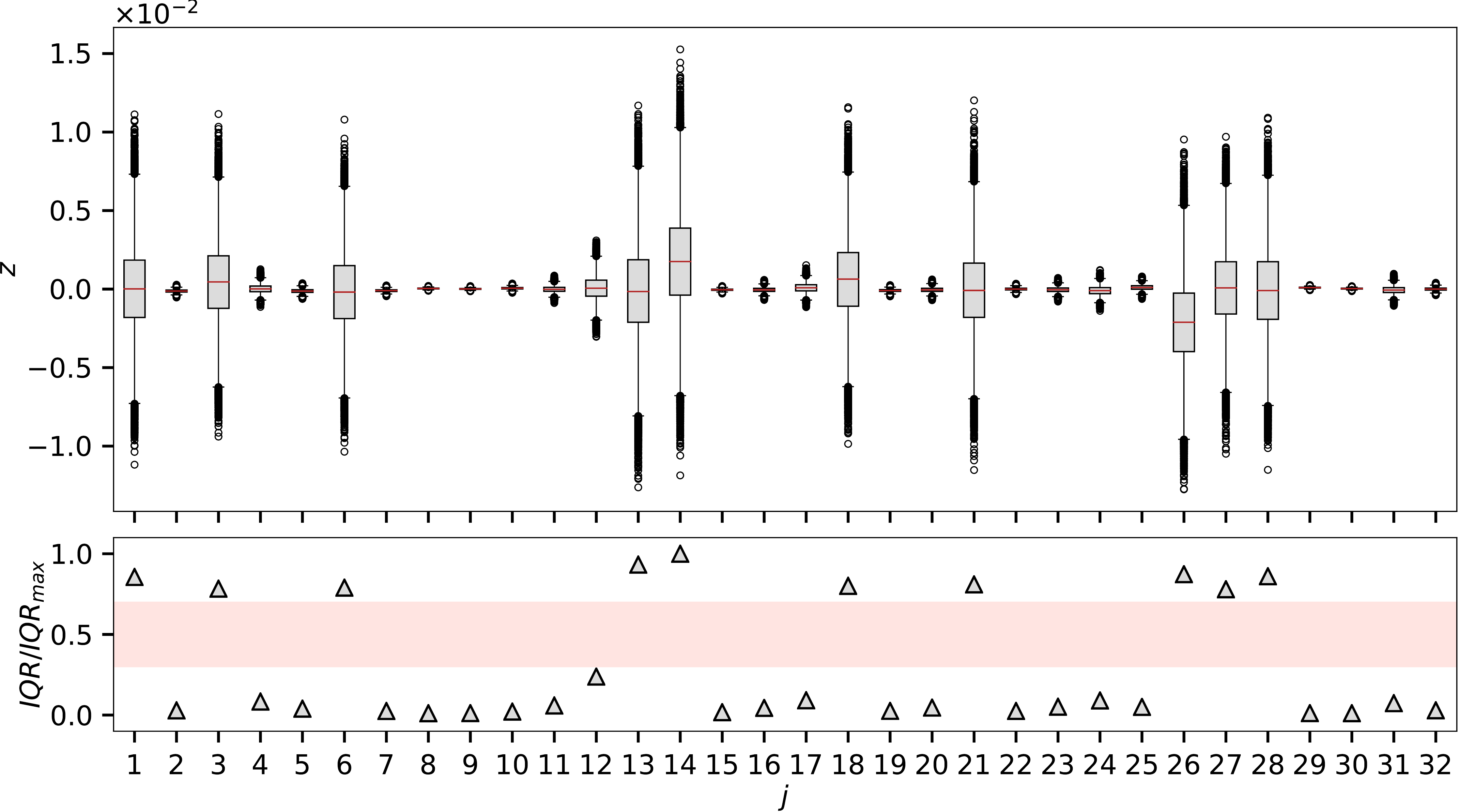}
  \caption{Active latent dimensions for the model with the lowest total loss~\eqref{eq:loss} ($\lambda=0.398$). Top: Box plots of each dimension of the latent space generated by passing all snapshots through the autoencoder and extracting the latent space. Bottom: Normalized $IQR$ for each dimension showing a gap between the $10$ active and remaining non-active dimensions.}
  \label{fig:IQR}
\end{figure}

Using the K-S dataset with $L=22$, the autoencoder was trained to find $\widehat{u}\left(x,t\right)$ with the total loss~\eqref{eq:loss} for $\lambda\in\left[10^{-4}, 2\right]$. The architecture of the encoder was $512 \rightarrow 256 \rightarrow 128 \rightarrow 64 \rightarrow 32$, corresponding to $N_{z}=32$. Figure~\ref{fig:summary} (a) presents $\mathcal{L}_{u}^{*}\lr{u_{\text{valid}},\widehat{u}_{\text{valid}}}$ for different values of $\lambda$ and shows a minimum at $\lambda=0.398$. Figure~\ref{fig:summary} (b) shows the reconstruction by the trained autoencoder using $\lambda=0.398$ on a snapshot from the validation set. More insight can be obtained by passing each snapshot through the trained network and extracting the latent dimensions corresponding to each snapshot. This process results in $N_{t}$ vectors (one for each snapshot), each of size $N_{z}$. 
%and provides information on which latent dimensions are the most / least active.
Figure~\ref{fig:IQR} shows that of the $32$ latent dimensions, only $10$ are consistently active. This is consistent with, but slightly larger than, the known dimension of the inertial manifold for the K-S equation~\cite{yang2012geometry,ding2016estimating}.
The interquartile range ($IQR$) can be used as an indicator of the variability of the latent dimensions. When normalized by the largest $IQR$ (the most active latent dimension) a very clear separation between ``active'' and ``non-active'' dimensions emerges as shown in the bottom of Figure~\ref{fig:IQR}. The largest $IQR$ of the remaining $22$ dimensions is $0.238$ of the most active dimension.

\begin{figure*}
  \subfloat[\label{fig:L22-basis}]{%
    \includegraphics[width=0.8\linewidth]{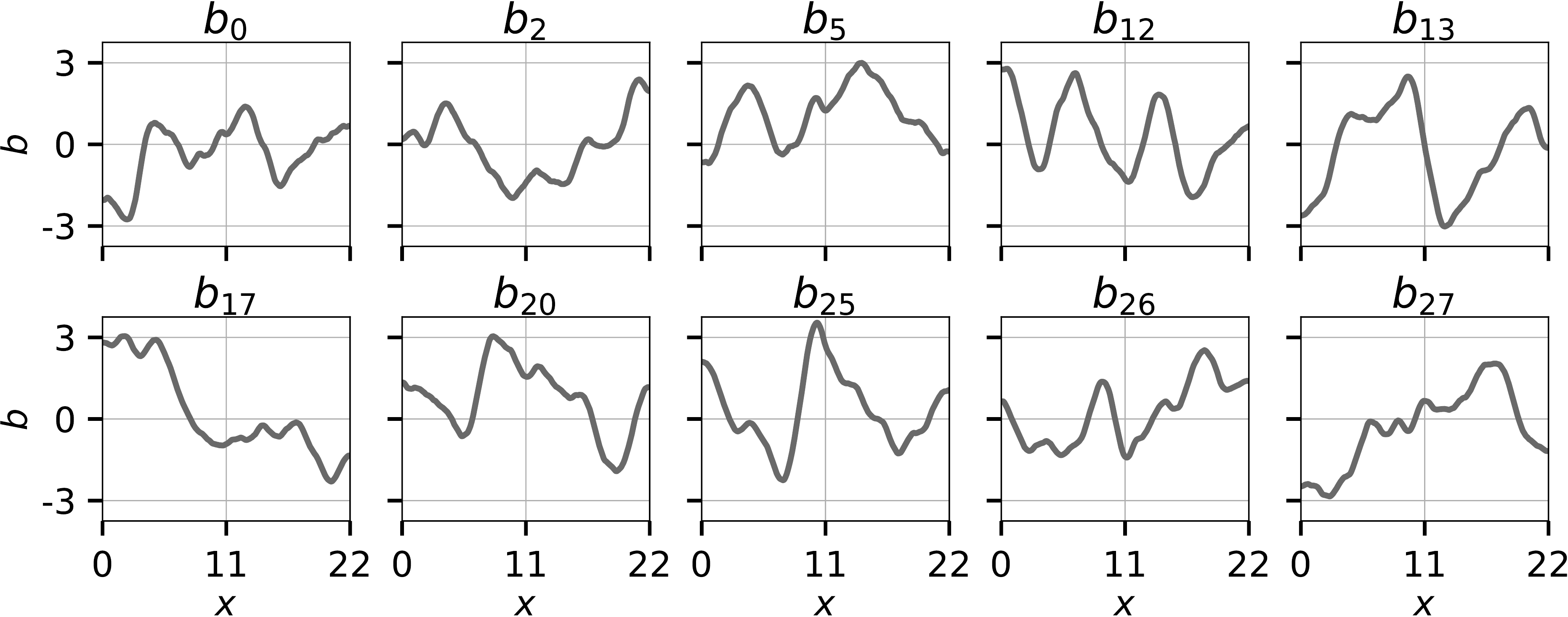}%
  }\hfill
  \subfloat[\label{fig:L22-nobasis}]{%
    \includegraphics[width=0.187\linewidth]{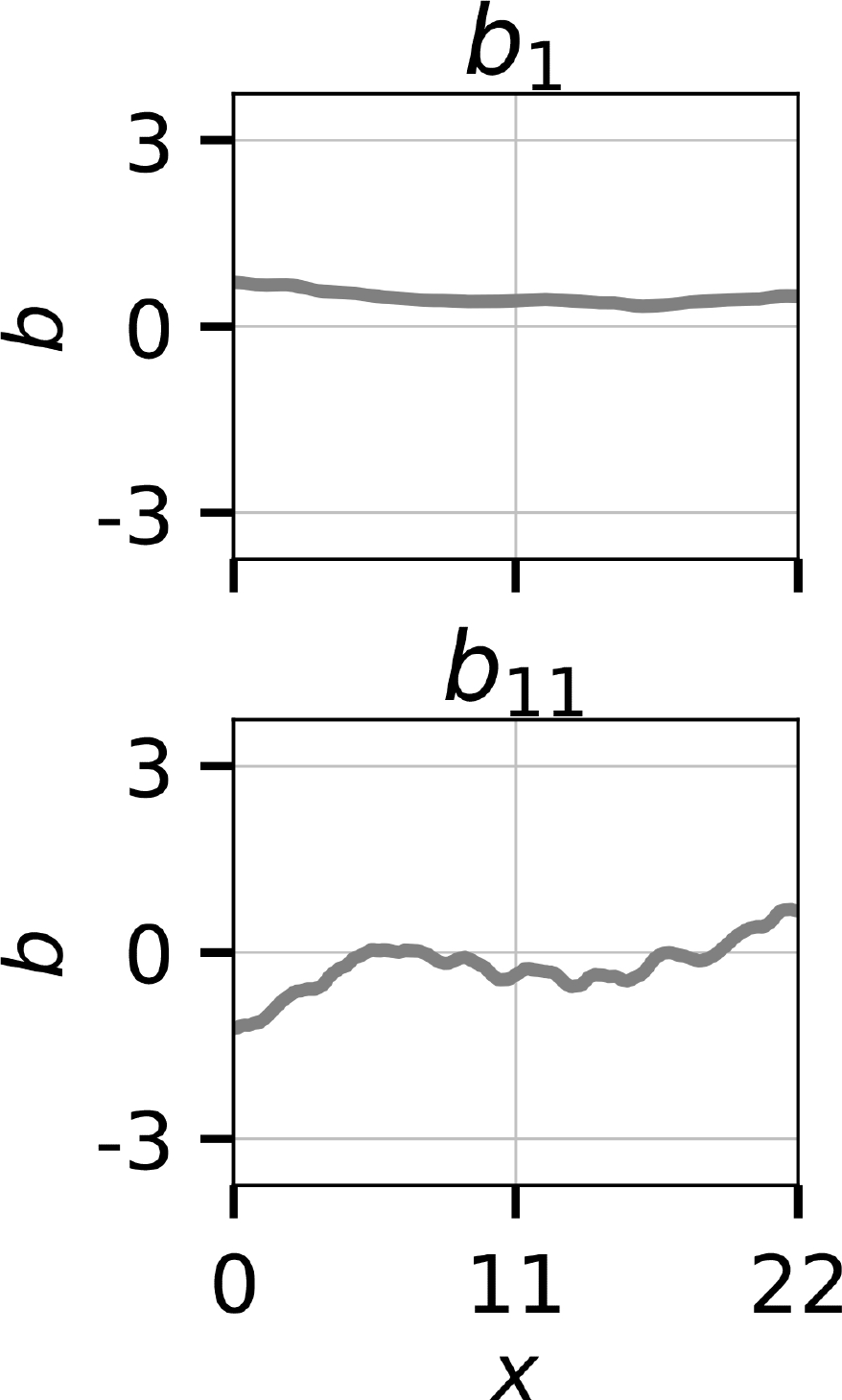}%
  }\\
  \subfloat[\label{fig:L22-basis_power}]{%
    \includegraphics[width=0.8\linewidth]{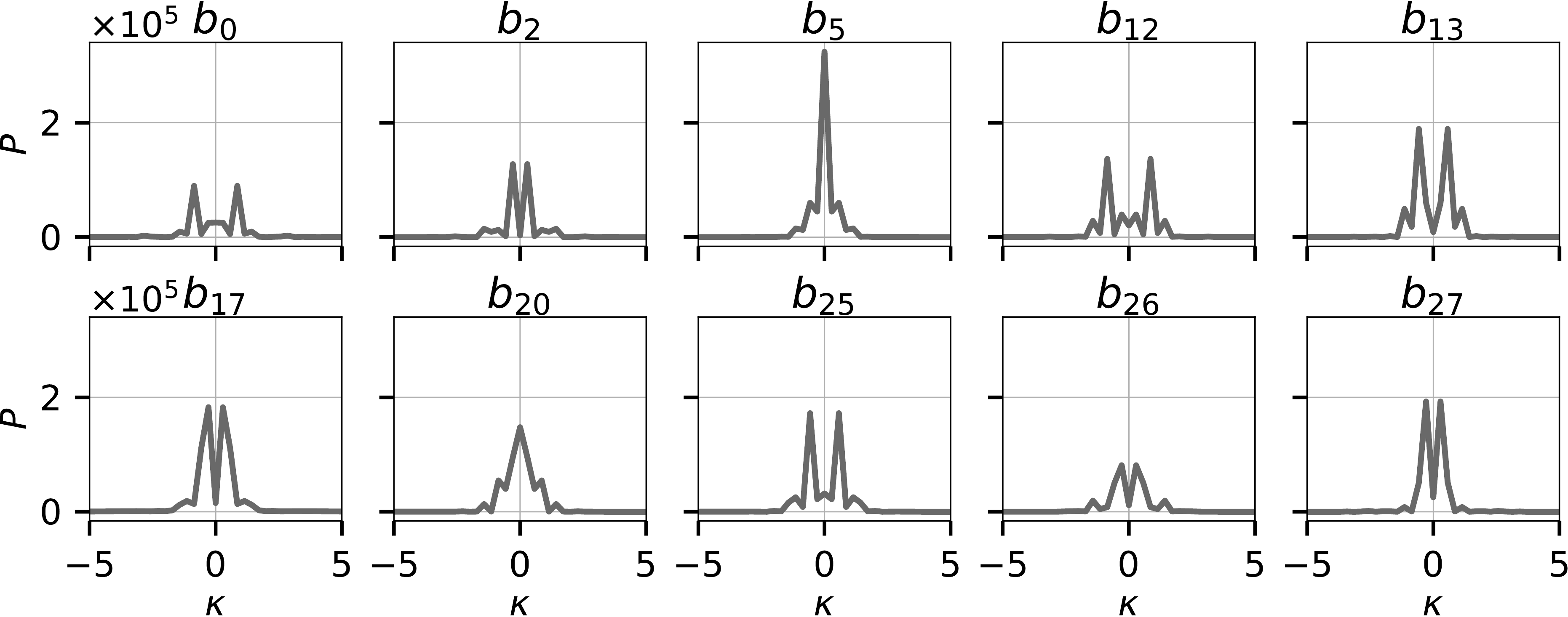}%
  }\hfill
  \subfloat[\label{fig:L22-nobasis_power}]{%
    \includegraphics[width=0.187\linewidth]{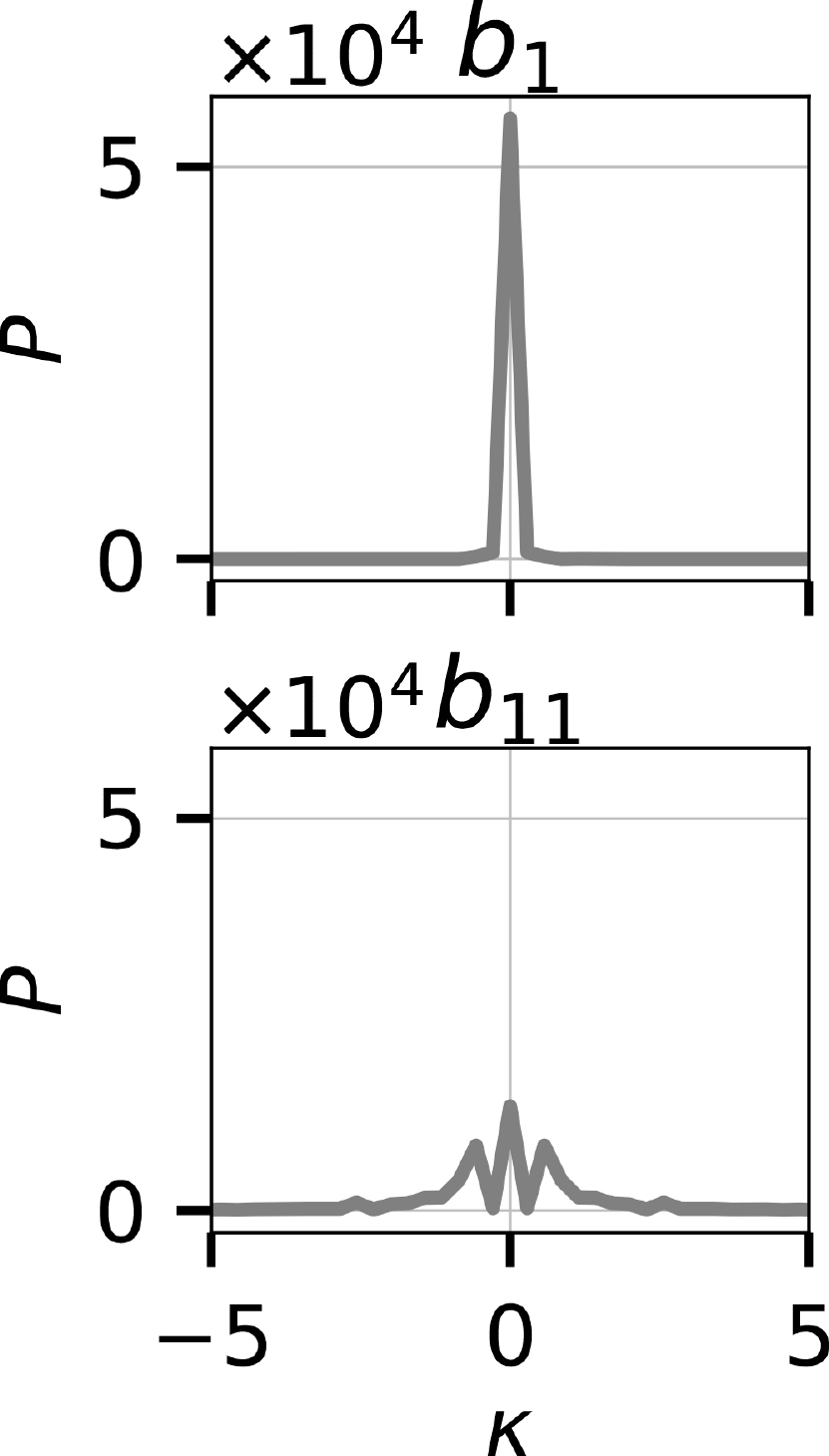}%
  }
  \caption{(a) The nonlinear basis for the $10-$ dimensional manifold found for the K-S equation with $L=22$. (b) Examples of the non-active dimensions. (c) Power spectrum of the basis for the active dimensions. (d) Power spectrum of two non-active dimensions. Wavenumbers in the power spectra have been truncated to highlight the peaks in the power spectra. There are no peaks for $\left|\kappa\right| > 5$.}
  \label{fig:basis}
\end{figure*}

\begin{figure}
  \centering
  \includegraphics[width=0.48\textwidth]{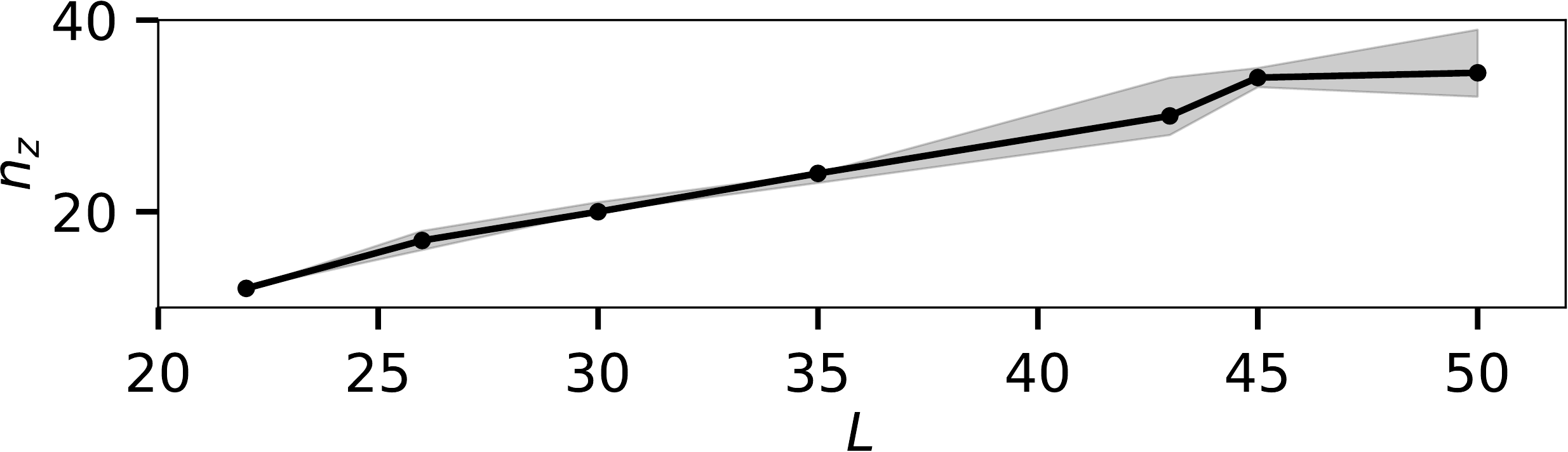}
  \caption{Scaling of the optimal number of latent dimensions with domain size $L$ for the K-S equation.}
  \label{fig:L_scaling}
\end{figure}

The trained autoencoder model also provides a way to develop a nonlinear basis for the learned manifold using the technique of activation maximization. 
%Starting from a random input drawn from a uniform distribution, $u_{\text{rand}}\sim\mathcal{U}\lr{0,1}$, to the trained autoencoder, gradient ascent can be performed on each dimension in the latent space to find the input that maximally activates each latent dimension. Regularization was used to smooth the resulting field during the maximization process. 
The idea behind this technique is to determine the input that maximizes the output of a specific node in the neural network. In the present work, we were interested in the inputs that would maximize each latent dimension. The input that maximized a given latent dimension was interpreted as a component of the basis of the low-dimensional manifold. The input that maximizes latent dimension $z_{j}$ is determined from,
\begin{align}
  b_{j} = \argmax_{v}S_{j}\lr{v}, \quad j = 1,\ldots, N_{z} \label{eq:argmax}
\end{align}
where 
\begin{align}
  S_{j} = z_{j}\lr{v} - \beta R_{TV}\lr{v}
\end{align}
and
\begin{align}
  R_{TV}\lr{v} = \frac{1}{N-1}\sum_{i=1}^{N-1}{\lr{v\lr{x_{i+1}} - v\lr{x_{i}}}^{2}}
\end{align}
is a regularization used to smooth the resulting field with the regularization parameter $\beta \geq 0$. The function evaluation $z_{j}\lr{v}$ corresponds to an evaluation of the encoder portion of the trained autoencoder with input $v$. Note that $v$ is not a temporal snapshot of the dataset, but instead represents an arbitrary input to the autoencoder. Gradient ascent was used to solve~\eqref{eq:argmax},
\begin{align}
  v_{j}^{\lr{l+1}} = v_{j}^{\lr{l}} + \gamma \frac{\partial S_{j}}{\partial v_{j}}.
\end{align}
The step size $\gamma$ was set to $1$ and the regularization strength $\beta$ was set to $3$. The maximization of each latent dimension was initialized from a random distribution in space where each point was drawn uniformly in $\left[0,1\right]$. The result is a basis for the reduced manifold. Figures~\ref{fig:L22-basis} and~\ref{fig:L22-basis_power} show the components of the basis and their power spectra, respectively. The power spectra clearly show that the components of the discovered basis consist of a handful of distinct modes. In contrast to the active dimensions, Figures~\ref{fig:L22-nobasis} and~\ref{fig:L22-nobasis_power} show that the inputs that maximize the non-active dimensions are constants near zero. 
 Finally, we compute the optimal number of latent dimensions for $L=\left[22, 26, 30, 35, 43, 45, 50\right]$ and repeat the experiment ten times for each value of $L$. Figure~\ref{fig:L_scaling} depicts the scaling of the number of latent dimensions with $L$, along with uncertainty bounds, and shows a nearly linear scaling, consistent with the scaling of the number of active modes~\cite{yang2009hyperbolicity}.
 %The scaling is observed to be linear~\cite{yang2009hyperbolicity}. 

\subsection{KdV Equation}
\begin{figure}
  \centering
  \includegraphics[width=0.48\textwidth]{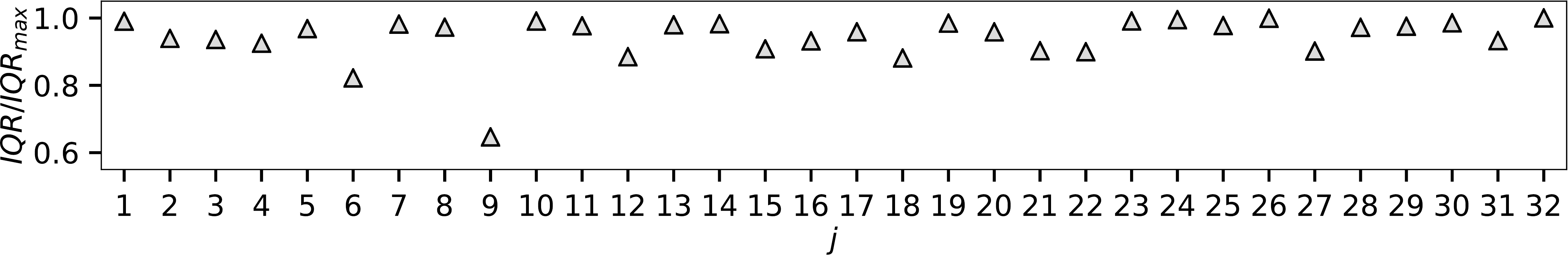}
  \caption{Active latent dimensions for the KdV equation at $\lambda=0$ visualized with the normalized $IQR$. The least active dimension is more than $60\%$ of the most active dimension indicating that all latent dimensions are required to reconstruct the solution.}
  \label{fig:kdv_boxplot.pdf}
\end{figure}

The latent space penalization enabled the discovery of a reduced basis for an equation that has an inertial manifold. In contrast to the K-S equation, the undamped KdV equation does not possess an inertial manifold and, moreover, the dynamics are truly infinite dimensional~\cite{ghidaglia1988weakly}. The undamped KdV equation therefore provides a test for the method in which the optimal latent space penalization is expected to be zero. Before training the autoencoder, the input was normalized by its maximum value. Following the same procedure as for the K-S equation, the optimal latent space regularization was determined to be $\lambda=0$ implying that the system does not possess an underlying finite-dimensional manifold. Figure~\ref{fig:summary} (c) presents a snapshot of a solution from the validation set and the corresponding prediction from the autoencoder at $\lambda=0$. The lack of a finite-dimensional manifold is further supported by Figure~\ref{fig:kdv_boxplot.pdf}, in which the $IQR$ for each snapshot of the dataset is visualized for each dimension in the latent space. There is no separation of the latent space dimensions into ``active'' and ``nonactive'' components.

\subsection{Damped KdV Equation}
\begin{figure*}
  \subfloat[\label{fig:shallow_min}]{%
    \includegraphics[width=0.50\linewidth]{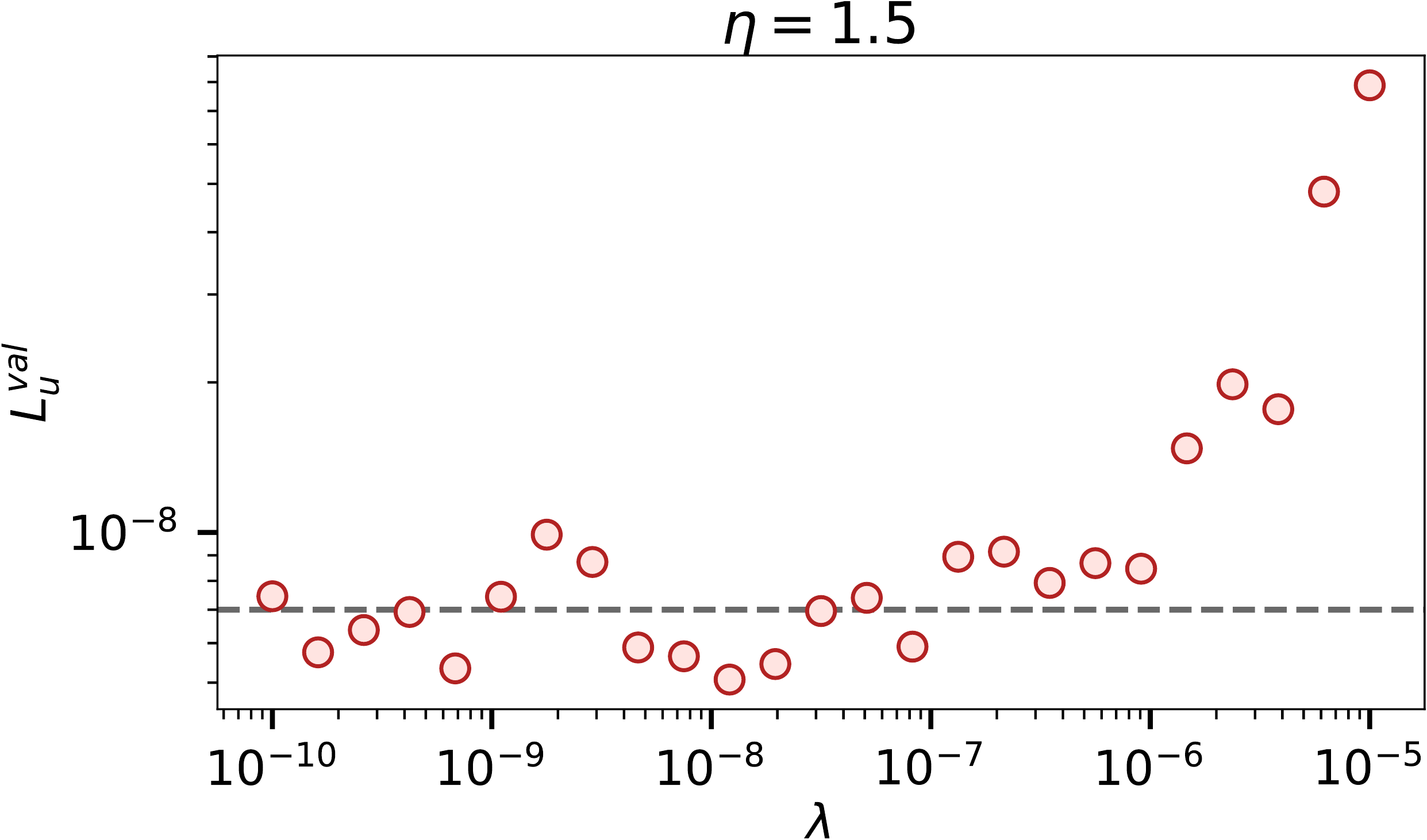}%
  }\hfill
  \subfloat[\label{fig:IQR_thresh}]{%
    \includegraphics[width=0.46\linewidth]{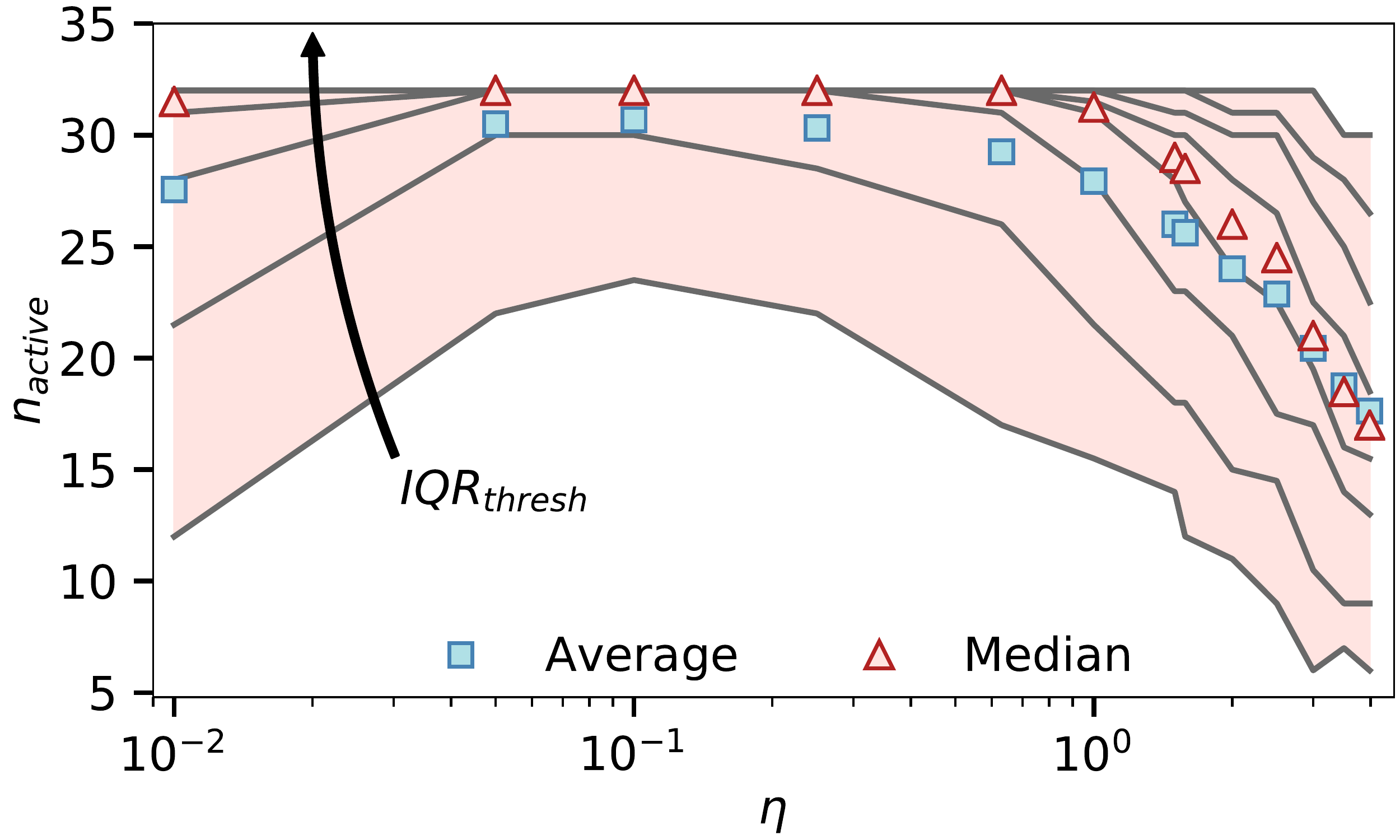}%
  }
  \caption{(a) Validation loss as a function of regularization parameter for the damped KdV equation with $\eta=1.5$. The dashed line is presented as a reference to highlight the shallow minimum / plateau. (b) The number of active dimensions as a function of damping parameter for different $IQR$ thresholds. The mean and median show the overall scaling. The paper presents the median.}
  \label{fig:shallow}
\end{figure*}

\begin{figure}
  \centering
  \includegraphics[width=0.48\textwidth]{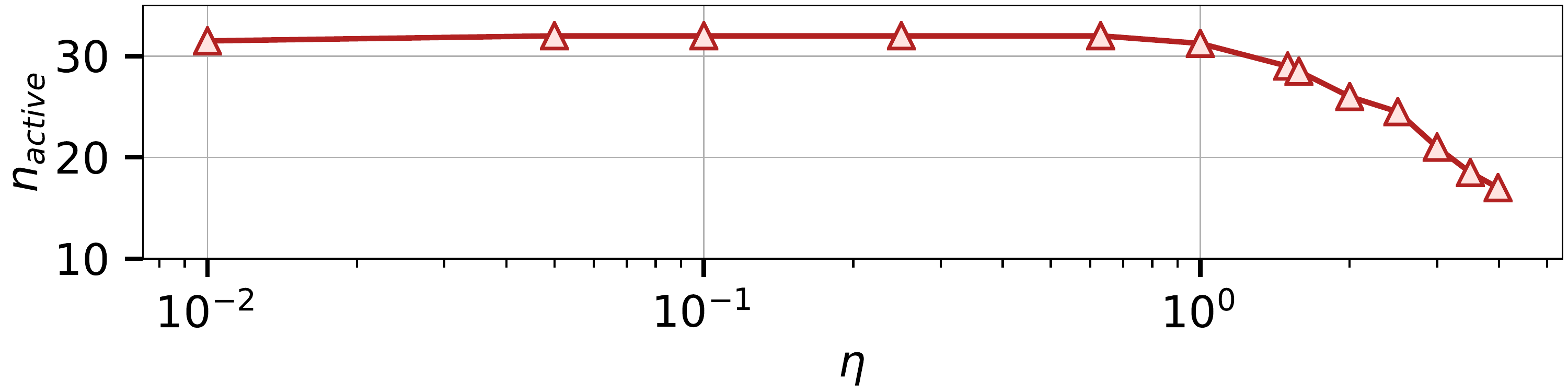}
  \caption{Active latent space dimensions vs. damping parameter $\eta$ for the damped KdV equation.}
  \label{fig:damped_kdv_dims}
\end{figure}

As a final test case, the damped KdV equation was studied over a range of damping coefficients, $\eta$ (see Table~\ref{tab:num_sim} in Appendix~\ref{app:num_sum}). The damped KdV equation possesses an inertial manifold and it therefore provides a fertile test ground for understanding the behavior of the latent space as the damping coefficient is varied. For each damping coefficient, the autoencoder with latent space penalization was trained across a number of regularization parameters. The number of active dimensions for each damped KdV equation had to be determined. However, the minima in the MSE loss, $L_{u}$ vs. regularization parameter, $\lambda$, on the validation sets were shallow. Figure~\ref{fig:shallow_min} presents a representative example for one particular model. Rather than extract a single minimum as was done in the K-S case, a different procedure was employed to provide a more robust result.
%Autoencoders with latent space penalization were trained for each damped KdV case as shown in Table~\ref{tab:AE_arch}. 
As before, the model with the minimum validation loss was extracted at each $\lambda$ and used for the analysis. Instead of retaining a single model at the minimum of the $L_{u}\lr{\lambda}$ curve, all models were retained in the shallow minimum. The latent dimensions for every snapshot of the dataset were computed for each of these models and the $IQR$ was again used to assess the activity of each latent dimension. The number of active dimensions was computed by counting the number of latent dimensions with $IQR$ above a certain threshold, $IQR_{\text{thresh}}$, for each model. To guard against sensitivity in the choice of $IQR_{\text{thresh}}$, a range of thresholds was tested between $0.1$ and $0.8$. This process results in an array representing the number of active dimensions for each value of $\eta$. The array contains the number of active dimensions for each model in the shallow minimum and each threshold value. The median number of dimensions over each model in the shallow minimum for each case and each threshold value was computed, which resulted in an array for each case that represents the number of active dimensions for each choice of threshold. As expected, the threshold value only shifts the number of active dimensions up and down, but the behavior of the number of active dimensions over $\eta$ for each threshold value remains the same. The final result is obtained by taking the median over the threshold values. Figure~\ref{fig:IQR_thresh} shows the variation of active dimensions as the threshold is increased.
The number of active latent dimensions decreases nearly monotonically with increasing damping coefficient as shown in Figure~\ref{fig:damped_kdv_dims}.

\section{Conclusions}
We introduced an autoencoder with latent space penalization applied to datasets generated from nonlinear partial differential equations in physics. The latent space penalization is used to restrict the dimensionality of the latent space to the minimal number of dimensions, thereby ensuring that the dynamics embodied in the dataset are captured by a low-dimensional manifold. In the case of the K-S equation, the optimal latent space dimension was consistent with the known dimensionality of the inertial manifold for bifurcation parameter $L=22$. We then determined a nonlinear basis for this manifold, which could in principle be used in a reduced order model. The KdV equation does not posses an inertial manifold and has truly infinite dimensional dynamics. In this case, we found optimal results without any latent space penalization, which is consistent with the known properties of the KdV equation. Finally, when applying this technique to the damped KdV equation, which once again has an inertial manifold, we found the beginnings of power law scaling for the number of active dimensions for sufficiently large damping coefficient. This technique opens interesting directions for determining reduced order models.
%that respect the known underlying physics.

\begin{acknowledgments}
The authors thank Michael Jolly and Robert Moser for insightful discussions. The computations in this paper were run on the FASRC Cannon cluster supported by the FAS Division of Science Research Computing Group at Harvard University.
\end{acknowledgments}

\bibliographystyle{apsrev}
\bibliography{refs}

\appendix

\section{Numerical Simulations}
\label{app:num_sum}
The one-dimensional Kuramoto-Sivashinsky (K-S) and Korteweg-de Vries (KdV) equations were solved using a pseudo-spectral Fourier discretization in space and an exponential fourth-order Runge-Kutta method~\cite{kassam2005fourth} in time. The spatial domain was discretized using $N$ points in physical space. The nonlinear terms were computed in physical space using the $2/3$ dealiasing rule. Table~\ref{tab:num_sim} presents the runs that were used to generate the figures in the paper.
\begin{table}
  \caption{\label{tab:num_sim} Details of the numerical simulations that were used to generate the datasets. The damping coefficient $\eta$ only applies to the damped KdV equation. $N$ represents the number of discrete points in the physical domain. The final integration time is $T$ and the constant time step is given by $\Delta t$.}
    \begin{ruledtabular}
      \begin{tabular}{lllllll}
      Case & Equation & Domain                   & $\eta$  & $N$     & $T$               & $\Delta t$ \\ \hline
      1    & K-S      & $\left[0, 22\right]$     & ---     & $512$   & $4\cdot 10^{4}$   & $0.125$    \\ 
      2    & K-S      & $\left[0, 22\right]$     & ---     & $1024$  & $4\cdot 10^{4}$   & $0.125$    \\ 
      3    & K-S      & $\left[0, 26\right]$     & ---     & $1024$  & $4\cdot 10^{4}$   & $0.125$    \\ 
      4    & K-S      & $\left[0, 30\right]$     & ---     & $1024$  & $4\cdot 10^{4}$   & $0.125$    \\ 
      5    & K-S      & $\left[0, 35\right]$     & ---     & $1024$  & $4\cdot 10^{4}$   & $0.125$    \\ 
      6    & K-S      & $\left[0, 43\right]$     & ---     & $1024$  & $4\cdot 10^{4}$   & $0.125$    \\ 
      7    & K-S      & $\left[0, 45\right]$     & ---     & $1024$  & $4\cdot 10^{4}$   & $0.125$    \\
      8    & K-S      & $\left[0, 50\right]$     & ---     & $1024$  & $4\cdot 10^{4}$   & $0.125$    \\
      9    & KdV      & $\left[-\pi, \pi\right]$ & $0$     & $512$   & $5\cdot 10^{-3}$  & $0.125$    \\
      10   & KdV      & $\left[-\pi, \pi\right]$ & $0.01$  & $512$   & $10^{-2}$         & $10^{-7}$ \\ 
      11   & KdV      & $\left[-\pi, \pi\right]$ & $0.05$  & $512$   & $10^{-2}$         & $10^{-7}$ \\
      12   & KdV      & $\left[-\pi, \pi\right]$ & $0.1$   & $512$   & $10^{-2}$         & $10^{-7}$ \\
      13   & KdV      & $\left[-\pi, \pi\right]$ & $0.25$  & $512$   & $10^{-2}$         & $10^{-7}$ \\
      14   & KdV      & $\left[-\pi, \pi\right]$ & $0.63$  & $512$   & $10^{-2}$         & $10^{-7}$ \\
      15   & KdV      & $\left[-\pi, \pi\right]$ & $1.0$   & $512$   & $10^{-2}$         & $10^{-7}$ \\
      16   & KdV      & $\left[-\pi, \pi\right]$ & $1.5$   & $512$   & $10^{-2}$         & $10^{-7}$ \\
      17   & KdV      & $\left[-\pi, \pi\right]$ & $1.58$  & $512$   & $10^{-2}$         & $10^{-7}$ \\
      18   & KdV      & $\left[-\pi, \pi\right]$ & $2.0$   & $512$   & $10^{-2}$         & $10^{-7}$ \\
      19   & KdV      & $\left[-\pi, \pi\right]$ & $2.5$   & $512$   & $10^{-2}$         & $10^{-7}$ \\
      20   & KdV      & $\left[-\pi, \pi\right]$ & $3.0$   & $512$   & $10^{-2}$         & $10^{-7}$ \\
      21   & KdV      & $\left[-\pi, \pi\right]$ & $3.5$   & $512$   & $10^{-2}$         & $10^{-7}$ \\
      22   & KdV      & $\left[-\pi, \pi\right]$ & $3.98$  & $512$   & $10^{-2}$         & $10^{-7}$ \\
      \end{tabular}
  \end{ruledtabular}
\end{table}

\section{Autoencoder Architectures}
\label{app:AE_arch}

\begin{table*}
  \caption{\label{tab:AE_arch} Parameters used for the trained autoencoders. The cases correspond to those in Table~\ref{tab:num_sim}. The Start Index column represents the snapshot index from which the training and validation sets were taken. Data before this index was not used.}
    \begin{ruledtabular}
      \begin{tabular}{llllll}
      Case & Architecture                                                        & Learning rate    & Batch size & Epochs             & Start Index \\ \hline
        1  & 
        \begin{tabular}{@{}c@{}} 
          (E): $512 \rightarrow 256 \rightarrow 128 \rightarrow 64 \rightarrow 32$  \\ 
          (D): $32 \rightarrow 64 \rightarrow 128 \rightarrow 256 \rightarrow 512$
        \end{tabular}
                                                                                 & $10^{-3}$        & $40$       & $3 \cdot 10^{4}$ & $5 \cdot 10^{4}$   \\ 
        2  & 
        \begin{tabular}{@{}c@{}} 
          (E): $1024 \rightarrow 256 \rightarrow 64$  \\ 
          (D): $64 \rightarrow 256 \rightarrow 1024$
        \end{tabular}
                                                                                 & $10^{-3}$        & $40$       & $3 \cdot 10^{4}$   & $5 \cdot 10^{4}$ \\ 
        3  & 
        \begin{tabular}{@{}c@{}} 
          (E): $1024 \rightarrow 256 \rightarrow 64$  \\ 
          (D): $64 \rightarrow 256 \rightarrow 1024$
        \end{tabular}
                                                                                 & $10^{-3}$        & $40$       & $3 \cdot 10^{4}$   & $5 \cdot 10^{4}$ \\ 
        4  & 
        \begin{tabular}{@{}c@{}} 
          (E): $1024 \rightarrow 256 \rightarrow 64$  \\ 
          (D): $64 \rightarrow 256 \rightarrow 1024$
        \end{tabular}
                                                                                 & $10^{-3}$        & $40$       & $3 \cdot 10^{4}$   & $5 \cdot 10^{4}$ \\ 
        5  & 
        \begin{tabular}{@{}c@{}} 
          (E): $1024 \rightarrow 256 \rightarrow 64$  \\ 
          (D): $64 \rightarrow 256 \rightarrow 1024$
        \end{tabular}
                                                                                 & $10^{-3}$        & $40$       & $3 \cdot 10^{4}$   & $5 \cdot 10^{4}$ \\ 
        6  & 
        \begin{tabular}{@{}c@{}} 
          (E): $1024 \rightarrow 256 \rightarrow 64$  \\ 
          (D): $64 \rightarrow 256 \rightarrow 1024$
        \end{tabular}
                                                                                 & $10^{-3}$        & $40$       & $3 \cdot 10^{4}$   & $5 \cdot 10^{4}$ \\ 
        7  & 
        \begin{tabular}{@{}c@{}} 
          (E): $1024 \rightarrow 256 \rightarrow 64$  \\ 
          (D): $64 \rightarrow 256 \rightarrow 1024$
        \end{tabular}
                                                                                 & $10^{-3}$        & $40$       & $3 \cdot 10^{4}$   & $5 \cdot 10^{4}$ \\
        8  & 
        \begin{tabular}{@{}c@{}} 
          (E): $1024 \rightarrow 256 \rightarrow 64$  \\ 
          (D): $64 \rightarrow 256 \rightarrow 1024$
        \end{tabular}
                                                                                 & $10^{-3}$        & $40$       & $3 \cdot 10^{4}$   & $5 \cdot 10^{4}$ \\
        9 ($\lambda = 0$) & 
        \begin{tabular}{@{}c@{}} 
          (E): $512 \rightarrow 256 \rightarrow 128 \rightarrow 64 \rightarrow 32$  \\ 
          (D): $32 \rightarrow 64 \rightarrow 128 \rightarrow 256 \rightarrow 512$
        \end{tabular}
                                                                                 & $5\cdot 10^{-3}$ & $100$      & $10^{5}$           & $10^{4}$         \\
        9 ($\lambda > 0$) & 
        \begin{tabular}{@{}c@{}} 
          (E): $512 \rightarrow 256 \rightarrow 128 \rightarrow 64 \rightarrow 32$  \\ 
          (D): $32 \rightarrow 64 \rightarrow 128 \rightarrow 256 \rightarrow 512$
        \end{tabular}
                                                                                 & $5\cdot 10^{-3}$ & $100$      & $10^{5}$           & $10^{4}$         \\
        10 & 
        \begin{tabular}{@{}c@{}} 
          (E): $512 \rightarrow 256 \rightarrow 128 \rightarrow 64 \rightarrow 32$  \\ 
          (D): $32 \rightarrow 64 \rightarrow 128 \rightarrow 256 \rightarrow 512$
        \end{tabular}
                                                                                 & $10^{-3}$        & $60$       & $5 \cdot 10^{4}$   & $10^{4}$         \\ 
        11 & 
        \begin{tabular}{@{}c@{}} 
          (E): $512 \rightarrow 256 \rightarrow 128 \rightarrow 64 \rightarrow 32$  \\ 
          (D): $32 \rightarrow 64 \rightarrow 128 \rightarrow 256 \rightarrow 512$
        \end{tabular}
                                                                                 & $10^{-3}$        & $60$       & $7.5 \cdot 10^{4}$ & $10^{4}$         \\
        12 & 
        \begin{tabular}{@{}c@{}} 
          (E): $512 \rightarrow 256 \rightarrow 128 \rightarrow 64 \rightarrow 32$  \\ 
          (D): $32 \rightarrow 64 \rightarrow 128 \rightarrow 256 \rightarrow 512$
        \end{tabular}
                                                                                 & $10^{-3}$  & $60$       & $7.5 \cdot 10^{4}$ & $10^{4}$         \\
        13 & 
        \begin{tabular}{@{}c@{}} 
          (E): $512 \rightarrow 256 \rightarrow 128 \rightarrow 64 \rightarrow 32$  \\ 
          (D): $32 \rightarrow 64 \rightarrow 128 \rightarrow 256 \rightarrow 512$
        \end{tabular}
                                                                                 & $10^{-3}$        & $60$       & $7.5 \cdot 10^{4}$ & $10^{4}$         \\
        14 & 
        \begin{tabular}{@{}c@{}} 
          (E): $512 \rightarrow 256 \rightarrow 128 \rightarrow 64 \rightarrow 32$  \\ 
          (D): $32 \rightarrow 64 \rightarrow 128 \rightarrow 256 \rightarrow 512$
        \end{tabular}
                                                                                 & $10^{-3}$        & $60$       & $7.5 \cdot 10^{4}$ & $10^{4}$         \\
        15 & 
        \begin{tabular}{@{}c@{}} 
          (E): $512 \rightarrow 256 \rightarrow 128 \rightarrow 64 \rightarrow 32$  \\ 
          (D): $32 \rightarrow 64 \rightarrow 128 \rightarrow 256 \rightarrow 512$
        \end{tabular}
                                                                                 & $10^{-3}$        & $60$       & $7.5 \cdot 10^{4}$ & $10^{4}$         \\
        16 & 
        \begin{tabular}{@{}c@{}} 
          (E): $512 \rightarrow 256 \rightarrow 128 \rightarrow 64 \rightarrow 32$  \\ 
          (D): $32 \rightarrow 64 \rightarrow 128 \rightarrow 256 \rightarrow 512$
        \end{tabular}
                                                                                 & $10^{-3}$        & $60$       & $7.5 \cdot 10^{4}$ & $10^{4}$         \\
        17 & 
        \begin{tabular}{@{}c@{}} 
          (E): $512 \rightarrow 256 \rightarrow 128 \rightarrow 64 \rightarrow 32$  \\ 
          (D): $32 \rightarrow 64 \rightarrow 128 \rightarrow 256 \rightarrow 512$
        \end{tabular}
                                                                                 & $10^{-3}$        & $60$       & $7.5 \cdot 10^{4}$ & $10^{4}$         \\
        18 & 
        \begin{tabular}{@{}c@{}} 
          (E): $512 \rightarrow 256 \rightarrow 128 \rightarrow 64 \rightarrow 32$  \\ 
          (D): $32 \rightarrow 64 \rightarrow 128 \rightarrow 256 \rightarrow 512$
        \end{tabular}
                                                                                 & $10^{-3}$        & $60$       & $7.5 \cdot 10^{4}$ & $10^{4}$         \\
        19 & 
        \begin{tabular}{@{}c@{}} 
          (E): $512 \rightarrow 256 \rightarrow 128 \rightarrow 64 \rightarrow 32$  \\ 
          (D): $32 \rightarrow 64 \rightarrow 128 \rightarrow 256 \rightarrow 512$
        \end{tabular}
                                                                                 & $10^{-3}$        & $60$       & $7.5 \cdot 10^{4}$ & $10^{4}$         \\
        20 & 
        \begin{tabular}{@{}c@{}} 
          (E): $512 \rightarrow 256 \rightarrow 128 \rightarrow 64 \rightarrow 32$  \\ 
          (D): $32 \rightarrow 64 \rightarrow 128 \rightarrow 256 \rightarrow 512$
        \end{tabular}
                                                                                 & $10^{-3}$        & $60$       & $7.5 \cdot 10^{4}$ & $10^{4}$         \\
        21 & 
        \begin{tabular}{@{}c@{}} 
          (E): $512 \rightarrow 256 \rightarrow 128 \rightarrow 64 \rightarrow 32$  \\ 
          (D): $32 \rightarrow 64 \rightarrow 128 \rightarrow 256 \rightarrow 512$
        \end{tabular}
                                                                                 & $10^{-3}$        & $60$       & $7.5 \cdot 10^{4}$ & $10^{4}$         \\
        22 & 
        \begin{tabular}{@{}c@{}} 
          (E): $512 \rightarrow 256 \rightarrow 128 \rightarrow 64 \rightarrow 32$  \\ 
          (D): $32 \rightarrow 64 \rightarrow 128 \rightarrow 256 \rightarrow 512$
        \end{tabular}
                                                                                 & $10^{-3}$        & $60$       & $7.5 \cdot 10^{4}$ & $10^{4}$         \\
      \end{tabular}
  \end{ruledtabular}
\end{table*}

\end{document}